\documentclass{jps-cp}
\usepackage{txfonts} 
\usepackage{bm}
\usepackage{braket}
\usepackage{color}

\title{Lattice QCD Study of the Nucleon-Charmonium Interaction}

\author{Takuya \textsc{Sugiura}$^{1}$, Yoichi \textsc{Ikeda}$^{1}$, and Noriyoshi \textsc{Ishii}$^{1}$}

\inst{$^{1}$Research Center for Nuclear Physics, Osaka University, 10-1 Mihogaoka, Ibaraki, Osaka, Japan}

\email{sugiura@rcnp.osaka-u.ac.jp}

\recdate{March 1, 2019}

\abst{The $J/\psi$-nucleon interaction is studied by lattice QCD
  calculations. At the leading order of the derivative expansion, the
  interaction consists of four terms: the central, the spin-spin, and
  two types of tensor forces. We determine these spin-dependent forces
  quantitatively by using the time-dependent HAL QCD method. We find
  that the spin-spin force is the main cause of the hyperfine
  splitting between the $J=1/2$ and the $J=3/2$ states, while the two
  tensor forces have much smaller effects on the S-wave scattering
  processes.}

\kword{lattice QCD, charmonium, spin-dependent forces}

\begin{document}
\maketitle

\section{Introduction}

One of the most important remaining problem in hadron physics is the
identification of exotic hadrons, which have more complex structure
than the conventional three-quark baryons or quark-antiquark mesons.
Recently, the LHCb collaboration has found peak structures in the
$J/\psi p$ invariant mass distribution from the weak decay $\Lambda_b
\to J/\psi K^- p$~\cite{LHCb1}.  As the $K^- p$ contribution alone
cannot describe the observed peaks~\cite{LHCb2}, they are expected to
be $uudc\bar{c}$ pentaquarks.  The Breit-Wigner fit shows that there
are two states, $P_c(4380)$ and $P_c(4450)$, with three acceptable
spin-parity combinations $(3/2^-, 5/2^+), (3/2^+, 5/2^-), (5/2^+,
3/2^-)$.

It is important to study the $J/\psi p$ scattering to search for $P_c$
states.  However, since experimental data in this channel is very
limited, phenomenological approaches may end up with large ambiguity.
To overcome this difficulty, we study the $J/\psi p$ scattering by
using lattice QCD simulations without requiring further experimental
data.  Specifically we employ the approach introduced by the HAL QCD
collaboration to calculate the $J/\psi p$ potential~\cite{HAL1, HAL2}.
Our previous report~\cite{Sugiura_Lattice} shows that we need to use
the time-dependent HAL QCD method to calculate the $J/\psi N$
potential properly.

The $J/\psi N$ interaction is expected to be dominated by the
exchanges of color-singlet gluon clusters, called the QCD van der
Waals forces~\cite{vanderWaals}.  The possibility of $P_c(4450)$ as a
deeply bound state of $\psi(2S)$ and $p$ has been discussed in this
framework~\cite{Polyakov1} .  In order to check this scenario
quantitatively, precise knowledge of the charmonium-nucleon
interaction is necessary~\cite{Polyakov2}.  In particular, the $J/\psi
N$ potential with spin-dependent forces is an essential key.  The
spin-dependent forces are omitted in many contexts, since they are
suppressed by $\mathcal{O}(1/m_c)$.  Almost no information of the
spin-dependent forces has been obtained so far.  The spin-dependent
potential is also important for a search for nucleus bound
$J/\psi$~\cite{Yokota}.
The aim of the present report is to perform a high-statistics lattice
QCD calculation to obtain the spin-dependent $J/\psi N$ potential from
the time-dependent HAL QCD method.

\section{Method}

In the time-dependent HAL QCD method, we start from the normalized
four-point correlation function for the $J/\psi N$ system,
\begin{align}
R(\bm{x}-\bm{y},t-t_0)
\equiv
\Braket{0| N(\bm{x},t) \psi(\bm{y}, t) \bar{\mathcal{J}}(t_0) | 0} / e^{-(m_N+m_\psi)(t-t_0)},
\end{align}
where $N(\bm{x},t)$, $\psi(\bm{y},t)$ and $\bar{\mathcal{J}}(t_0)$
denote local interpolating operators for proton and $J/\psi$, and the
corresponding wall-source operator.  In actual lattice QCD
calculations, the exponential factors involving the proton mass $m_N$
and the $J/\psi$ mass $m_\psi$ is replaced by the corresponding
single-hadron correlation functions.  The $J/\psi N$ non-local and
energy-independent potential $U(\bm{r}, \bm{r}^\prime)$ satisfies the
following time-dependent Schr\"odinger-like equation:
\begin{align}
\left( 
\frac{\bm{\nabla}^2}{2\mu}
+ \frac{\partial}{\partial t}
\right)
R(\bm{r},t)
=
\int d^3\bm{r}^\prime
U(\bm{r}, \bm{r}^\prime)
R(\bm{r}^\prime, t),
\end{align}
where $\mu=1/(1/m_N+1/m_\psi)$ is the reduced mass.  On the left-hand
side, small corrections of $\mathcal{O}(k^4)$ ( $k$ is related to the
center of mass energy W as $W=\sqrt{k^2+m_N^2}+\sqrt{k^2+m_\psi^2}$ )
are neglected, as we have confirmed that the corrections are
negligibly small.
By taking the leading-order terms of the derivative expansion, the
non-local potential is approximated by a local one, $U(\bm{r},
\bm{r}^\prime)\simeq V(r)\delta^3(\bm{r}-\bm{r}^\prime)$.  Here $V(r)$
has spin indices of $p$ and $J/\psi$.  Our next task is to derive the
spin structure of $V(r)$.

The nucleon spin operator is related to the Pauli matrices as
$\bm{\sigma}/2$, whereas the $J/\psi$ spin operator corresponds to the
spin-1 equivalents of the Pauli matrices,
\begin{align}
\Sigma_1 = 
	\begin{pmatrix}
		0 & 0 & 0  \\  0 & 0 & -i  \\  0 & +i & 0
	\end{pmatrix},
\quad
\Sigma_2 = 
	\begin{pmatrix}
		0 & 0 & +i  \\  0 & 0 & 0  \\  -i & 0 & 0
	\end{pmatrix},
\quad
\Sigma_3 = 
	\begin{pmatrix}
		0 & -i & 0  \\  +i & 0 & 0  \\  0 & 0 & 0
	\end{pmatrix},
\end{align}
which satisfy the commutation relations $\left[ \Sigma_i, \Sigma_j
  \right]=i\epsilon_{ijk}\Sigma_k$.  The total spin operator $\bm{S}$
is thus given as $\bm{S}=\bm{\sigma}/2+\bm{\Sigma}$, which has
eigenvalues $1/2$ and $3/2$.  We follow the arguments of
Okubo~and~Marshak~\cite{OkuboMarshak} to derive the general form of
the $J/\psi N$ 2-body interaction at the leading order of the
derivative expansion.  We require that the potential satisfy the
following six conditions: energy-momentum conservation, Galilei
covariance, isospin invariance, total angular momentum conservation,
Parity invariance, and time-reversal invariance.  Then we find that
the leading-order potential in the center of mass frame must have the
form
\begin{align}
\label{eq:JpsiN_potential_general}
V(r) =
V_0(r)
+
V_s(r)	  \bm{\sigma} \cdot \bm{\Sigma}
+
V_{T1}(r) S_{12}
+ 
V_{T2}(r) T_{12}
\end{align}
where the relative coordinate is denoted by $\bm{r}=r\,\hat{\bm{r}}$,
and $S_{12} = 3 \left( \hat{\bm{r}} \cdot \bm{\sigma} \right) \left(
\hat{\bm{r}} \cdot \bm{\Sigma} \right) - \bm{\sigma} \cdot
\bm{\Sigma}$ and $T_{12} = 3 \left( \hat{\bm{r}} \cdot \bm{\Sigma}
\right)^2 - \bm{\Sigma}^2$ are the two tensor operators of the $J/\psi
N$ interaction.  The convergence of the derivative expansion must be
checked numerically.

There are five $J/\psi N$ partial waves with total angular momentum
$J=1/2$ or $3/2$: $^2S_{1/2}$, $^4D_{1/2}$, $^4S_{3/2}$, $^2D_{3/2}$,
and $^4D_{3/2}$, in the standard notation $^{2s+1}l_J$.
The four forces $V_0(r)$, $V_s(r)$, $V_{T1}(r)$ and $V_{T2}(r)$ are
determined from the wave functions for four of the five states.
In
general the $J/\psi N$ system with $J^P=1/2^-$ couples to the $\eta_c N$
system with the same quantum numbers; however we assume this transition
is negligible due to the heavy charm quark mass.

\section{Simulation Setup}

We employ the $2+1$ flavor QCD gauge configurations on a $32^3\times
64$ lattice generated by the PACS-CS collaboration~\cite{PACS-CS}.
They are generated with the renormalization group improved gauge
action at $\beta=1.90$ and the non-perturbatively $\mathcal{O}(a)$
improved Wilson quark action at $c_{SW}=1.715$, which correspond to
lattice spacing of $a=0.0907(13) \mathrm{fm}$ and the spatial volume
of $L^3=(2.9\mathrm{fm})^3$.  The hopping parameters are taken to be
$\kappa_{ud}=0.13700$ and $\kappa_{s}=0.13640$.  For the charm quark
we employ the Tsukuba-type relativistic heavy quark (RHQ) action to
remove the leading and next-to-leading order cutoff
errors~\cite{RHQ1}.  We use the same RHQ action parameters as in
Ref.~\cite{RHQ2}.
The periodic boundary condition is imposed in the spatial directions,
while the Dirichlet boundary condition is imposed in the temporal
direction at $t-t_0=32a$ to prevent inverse propagation.  To improve
statistics, we take average over $32$ different source positions
$t_0$.
In this setup, the masses of pion, proton, and $J/\psi$ are found to
be $m_\pi = 700(2) \mathrm{MeV}$, $m_N=1585(16) \mathrm{MeV}$, and
$m_\psi=3139(12) \mathrm{MeV}$, respectively.

\section{Results}

\begin{figure}[b]
	\begin{minipage}{0.5\hsize}
	\centering
	\includegraphics[width=\hsize,bb=0 0 792 612]{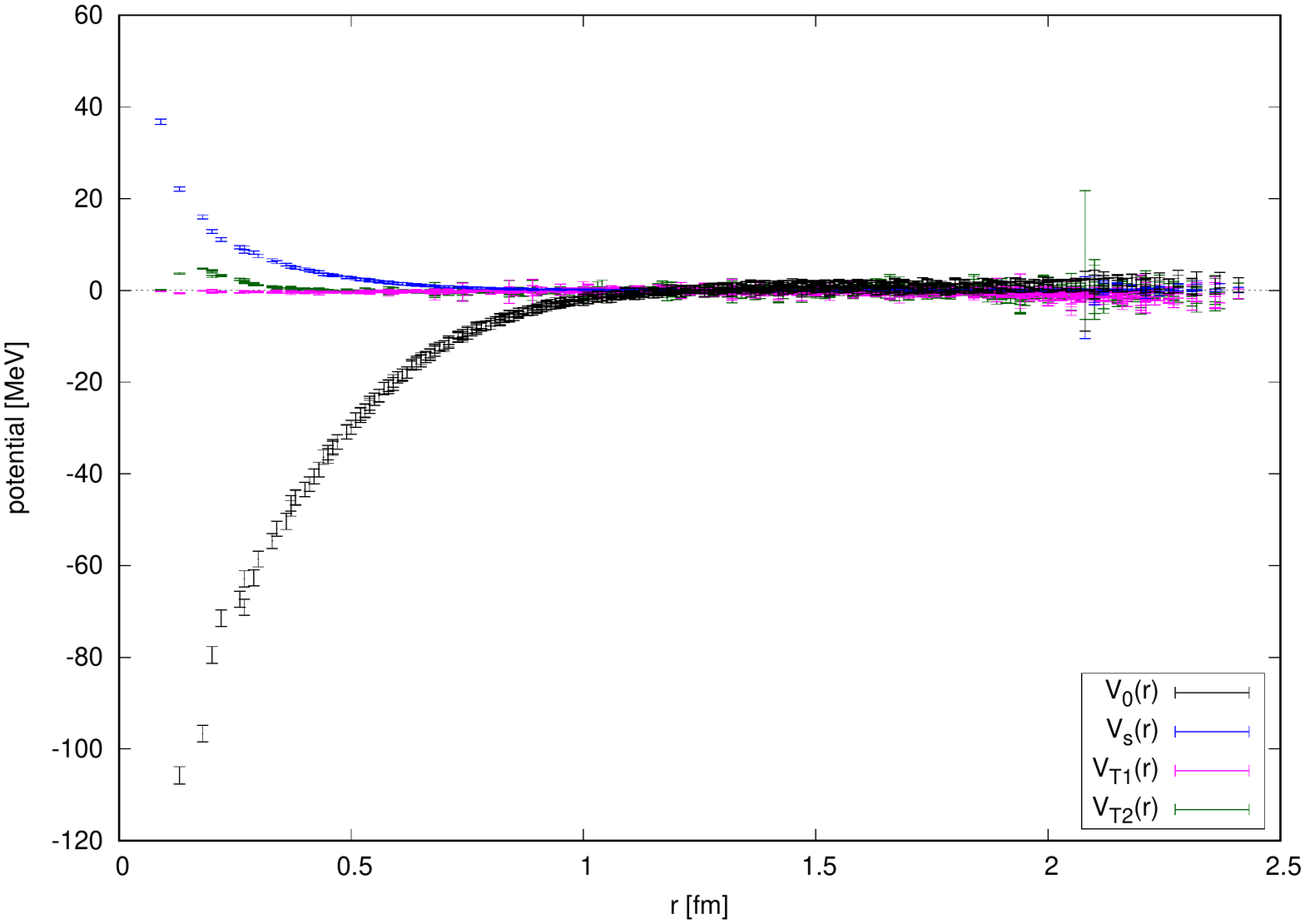}
	\end{minipage}
	\begin{minipage}{0.5\hsize}
	\centering
	\includegraphics[width=\hsize,bb=0 0 792 612]{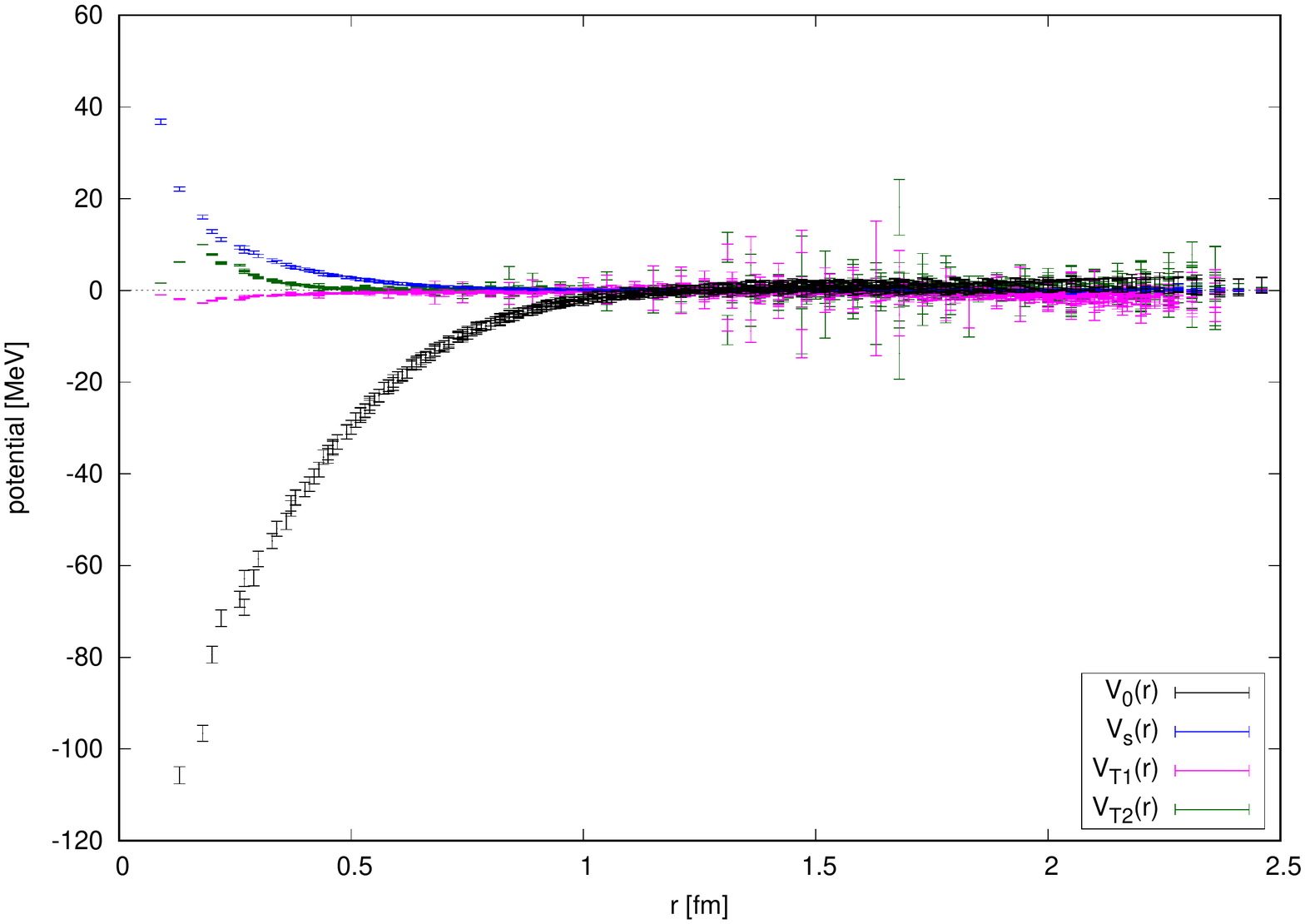}
	\end{minipage}
\caption{\label{fig:pot_spin} The $J/\psi N$ potentials $V_0(r)$,
  $V_s(r)$, $V_{T1}(r)$, and $V_{T2}(r)$ at $t-t_0=15a$. The left
  panel shows the potentials extracted from the states $({}^2S_{1/2},
  {}^4S_{3/2}, {}^2D_{3/2}, {}^4D_{3/2})$, while the right panel is
  from $({}^2S_{1/2}, {}^4D_{1/2}, {}^4S_{3/2}, {}^4D_{3/2})$.  }
\end{figure}

In Fig.~\ref{fig:pot_spin}, we show the $J/\psi N$ potentials
$V_0(r)$, $V_s(r)$, $V_{T1}(r)$, and $V_{T2}(r)$ at $t-t_0=15a$.  The
left panel shows the potentials extracted from the combination of
states $({}^2S_{1/2}, {}^4S_{3/2}, {}^2D_{3/2}, {}^4D_{3/2})$, while
the right panel is from $({}^2S_{1/2}, {}^4D_{1/2}, {}^4S_{3/2},
{}^4D_{3/2})$.  The difference of the two implies higher-order
contributions of the derivative expansion.  We see that $V_0(r)<0$ and
$V_s(r)>0$ for all $r$, and there is no visible difference in $V_0(r)$
and $V_s(r)$ between the two panels.  The strength of the tensor
forces $V_{T1}(r)$ and $V_{T2}(r)$ are relatively smaller compared to
$V_0(r)$ or $V_s(r)$.  Also, we see significant differences in the
strength of the tensor forces between the two panels in the small $r$
region.  In our simulation setup, the D-wave signals are very small,
and the tensor forces tend to contain larger ambiguity than the
central and the spin-spin forces.  It will be interesting to study the
tensor forces with better precision by using a different source
operator that couples to the D-wave states.
Note that, in the present analysis we assume that the
  wave functions are the irreducible representations of rotational
  SO(3) symmetry.  As a result, the potentials involve large
  ambiguities at points $(x,y,z)=(\pm n, \pm n, \pm n)$ ($n \in
  \mathbb{Z}$), so that we have removed them in
  Fig.~\ref{fig:pot_spin}.  The large erorrs at large distances will
  be due to the same reason. 
The temporal separation $t-t_0$ has to be large enough to achieve the
elastic-state saturation.  A necessary condition for this is that the
potentials are not changed by small variation of $t$.  We have
confirmed that the latter condition is satisfied at $t-t_0=15a$, so
that the elastic-state saturation is expected to be achieved.

\begin{figure}
	\begin{minipage}{0.5\hsize}
	\centering
	\includegraphics[width=\hsize,bb=0 0 792 612]{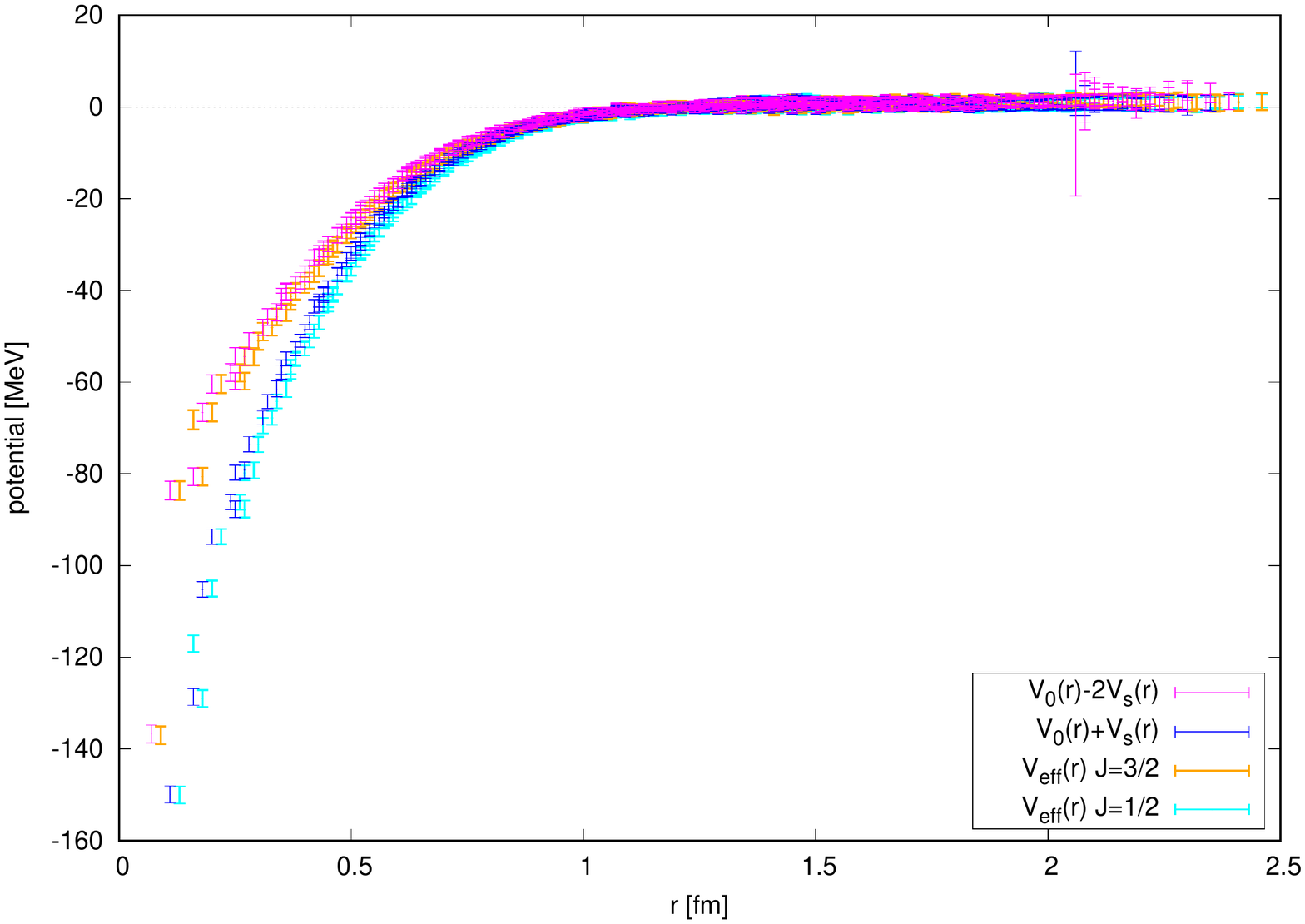}
	\end{minipage}
	\begin{minipage}{0.5\hsize}
	\centering
	\includegraphics[width=\hsize,bb=0 0 792 612]{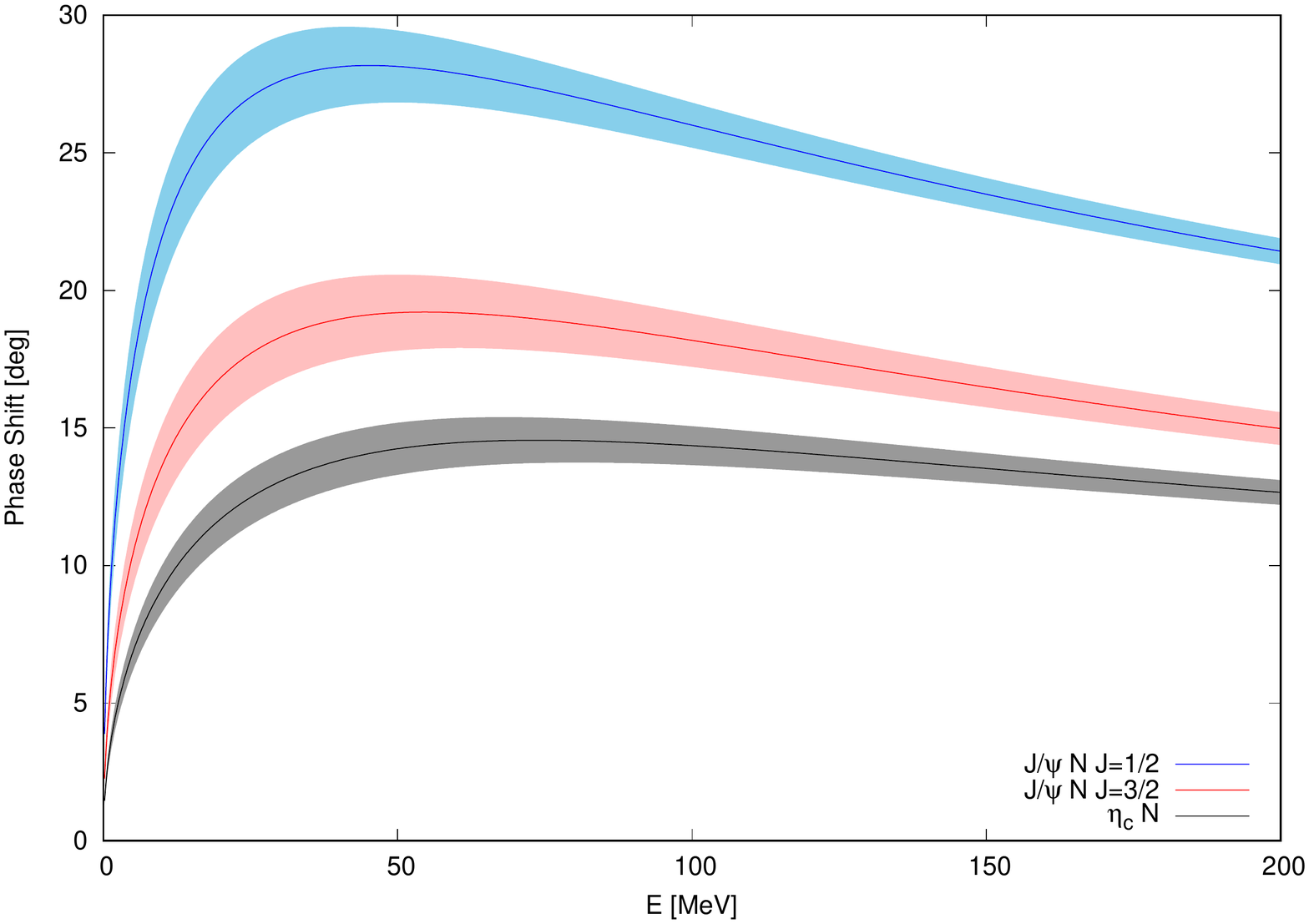}
	\end{minipage}
\caption{\label{fig:Veff_and_phaseshift} (Left) Linear combinations
  $V_0(r)-2V_s(r)$ and $V_0(r)+V_s(r)$, compared to the corresponding
  effective central potentials $V_{\text{eff}}^{(J=1/2)}(r)$ and
  $V_{\text{eff}}^{(J=3/2)}(r)$.  Note that $V_{\text{eff}}(r)$ are
  shifted in the $r$ direction by $-0.02$ for visibility.  \quad
  (Right) The scattering phase shift from the S-wave effective central
  potentials for $J/\psi N$ ($J=1/2$), $J/\psi N$ ($J=3/2$), and
  $\eta_c N$.  }
\end{figure}

By neglecting the $S-D$ mixing contributions from the tensor forces,
the S-wave diagonal matrix elements of the potential in
Eq.~\eqref{eq:JpsiN_potential_general} are
\begin{align}
\Braket{ {}^2S_{1/2} | V(r) | {}^2S_{1/2} } &= V_0(r) - 2 V_s(r), \\
\Braket{ {}^4S_{3/2} | V(r) | {}^4S_{3/2} } &= V_0(r) + V_s(r).
\end{align}
In Fig.~\ref{fig:Veff_and_phaseshift}(Left), we show these diagonal
matrix elements together with the S-wave effective central potentials
(for the definition of the effective central potentials, see
Ref.~\cite{Sugiura_Lattice}.).  We see that each diagonal matrix
element coincides with great precision with the corresponding
effective central potential.  It implies that the hyperfine splitting
of the $J/\psi N$ system between $J=1/2$ and $J=3/2$ is dominated by
the spin-spin force, and the contribution from the tensor forces are
much smaller.

To solve the Schr\"odinger equation for the scattering phase shift, we
neglect the small tensor force contributions, so that the S-wave
scattering is described by the effective central potentials with good
precision.  Shown in Fig.~\ref{fig:Veff_and_phaseshift}(Right) are the
calculated phase shifts.  For comparison we also include the
corresponding result for the $\eta_c N$ scattering.  We see that there
is no two-body bound state below the charmonium-nucleon production
thresholds. The scattering length $a$ and the effective range $r$ are
given as $a(J/\psi N J=1/2)=0.656(71) \mathrm{fm}$, $r(J/\psi N
J=1/2)=1.105(16) \mathrm{fm}$, $a(J/\psi N J=3/2)=0.380(48)
\mathrm{fm}$, $r(J/\psi N J=3/2)=1.476(39) \mathrm{fm}$, $a(J/\psi N
J=1/2)=0.246(26) \mathrm{fm}$, and $r(J/\psi N J=1/2)=1.703(45)
\mathrm{fm}$.

\vspace{-10pt}
\vspace{-15pt}

\section{Conclusions}

We have studied the $J/\psi N$ scattering by the time-dependent HAL
QCD method.  We find that the $J/\psi N$ interaction at the leading
order of the derivative expansion consists of the central, the
spin-spin, and the two tensor forces.  The spin-spin force strengthen
the attraction for $J=1/2$ and weaken the attraction for $J=3/2$,
causing the hyperfine splitting between the two states. The tensor
forces are much weaker than the central and the spin-spin forces.

\section{Acknowledgements}

This study is supported by Japan Society for the Promotion of Science
KAKENHI Grands No.~JP25400244 and by Ministry of Education, Culture,
Sports, Science and Technology as ``Priority Issue on Post-K
computer'' (Elucidation of the Fundamental Laws and Evolution of the
Universe) and Joint Institute for Computational Fundamental
Science. We thank PACS-CS Collaboration~\cite{PACS-CS} and
ILDG/JLDG~\cite{ILDG_JLDG} for providing the gauge configurations. The
numerical calculations have been performed on Oakforest-PACS at the
University of Tokyo and OCTOPUS in Osaka University.  The lattice QCD
code is partly based on Bridge++~\cite{Bridge}.

\end{document}